\documentclass[manuscript]{acmart}

\usepackage{colortbl}
\usepackage{url}
\usepackage[normalem]{ulem}
\usepackage{subfig}  
\usepackage{enumitem}

\AtBeginDocument{%
  }

\begin{document}

\title{Reflection at Design Actualization (RDA) : A Tool and Process For Research Through Game Design}

\author{Prabhav Bhatnagar}
\orcid{0000-0001-7196-5254}
\email{prabhav.bhatnagar@aalto.fi}
\affiliation{%
  \institution{Aalto University}
  \city{Espoo}
  \country{Finland}}

\author{Jianheng He}
\orcid{0009-0006-2297-1619}
\email{jianheng.he.24@ucl.ac.uk}
\affiliation{%
  \institution{University College London}
  \city{London}
  \country{United Kingdom}}

  \author{Shamit Ahmed}
\orcid{0009-0002-7462-9760}
\email{shamit.ahmed@aalto.fi}
\affiliation{%
  \institution{Aalto University}
  \city{Espoo}
  \country{Finland}}

\author{Andrés Lucero}
\orcid{0000-0002-7176-2884}
\email{andres.lucero@aalto.fi}
\affiliation{%
  \institution{Aalto University}
  \city{Espoo}
  \country{Finland}}

\author{Perttu Hämäläinen}
\orcid{0000-0001-7764-3459}
\email{perttu.hamalainen@aalto.fi}
\affiliation{%
  \institution{Aalto University}
  \city{Espoo}
  \country{Finland}}

\renewcommand{\shortauthors}{Prabhav Bhatnagar, Jianheng He, Shamit Ahmed, Andrés Lucero \& Perttu Hämäläinen}

\begin{abstract}

There is a growing interest in researching game design processes, artifacts and culture through active game design. Tools and processes to support these attempts are limited, especially in terms of a) capturing smaller design decisions where rich tacit information is often situated, and b) visually tracking the project's growth and evolution. To address this gap, we present Reflection at Design Actualization~(RDA), an open source tool and process for collecting granular reflections at playtesting moments and automatically recording the playtests, bringing reflection and data collection closer to the point where design decisions concretize. Three researchers engaged with and evaluated RDA in three varied game development projects, adhering to the principles of autobiographical design. We illustrate the designer experience with RDA through three themes, namely, designer-routine compromise, designer-researcher persona consolidation, and mirror effect of RDA. We further discuss the tool's challenges and share each designer's personal experience as case studies. 

\end{abstract}

\ccsdesc[500]{Software and its engineering~Interactive games}
\ccsdesc[500]{Applied computing~Computer games}
\ccsdesc[500]{Human-centered computing~Interaction design process and methods}

\keywords{game design,video games, tool development, autobiographical design, research-through-design, research-through-game-design}


\setlength{\belowcaptionskip}{5pt}
\setlength{\abovecaptionskip}{7pt}

  \begin{teaserfigure}
    \centering
    \includegraphics[width = 0.95 \linewidth]{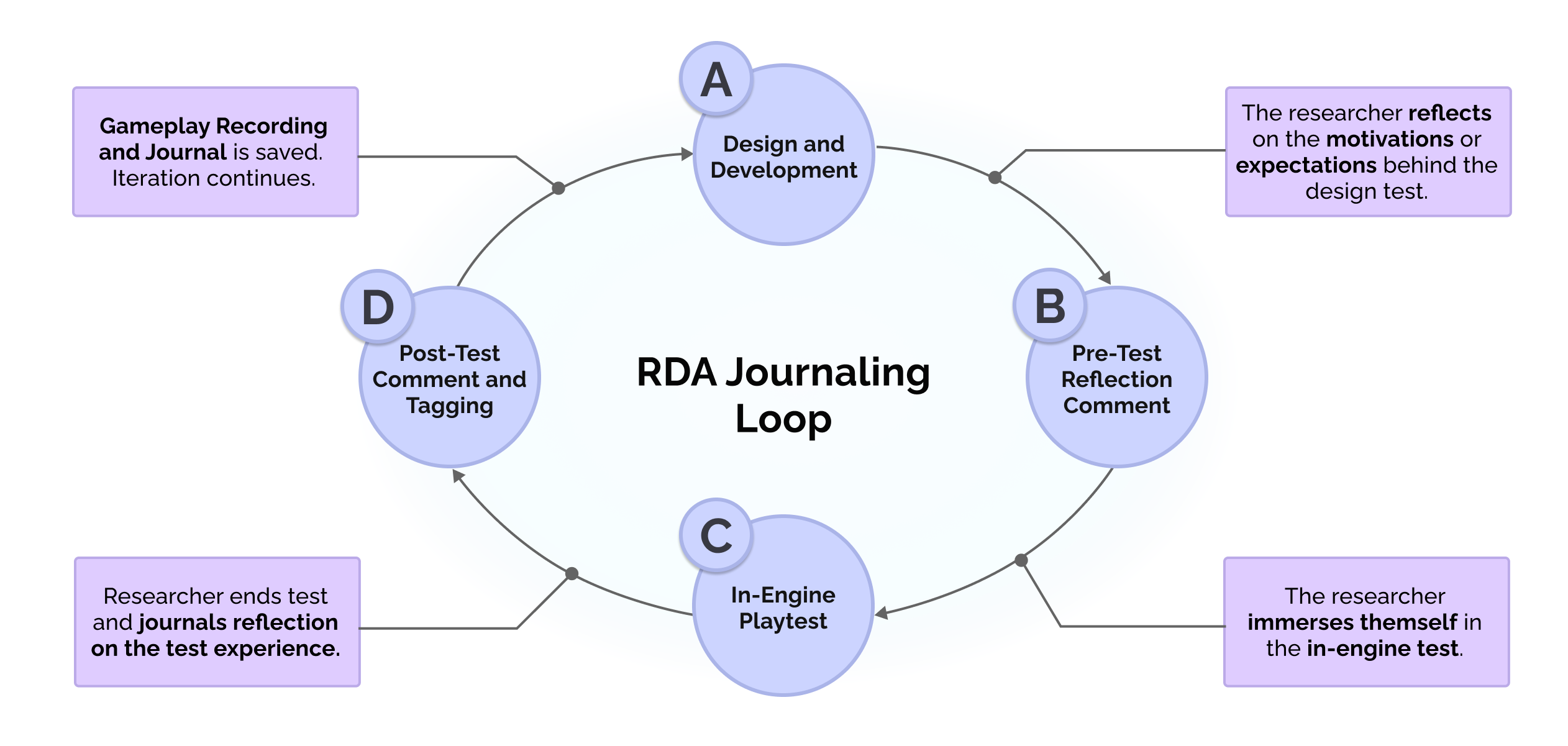}
    \caption{The journaling loop of the proposed RDA tool interleaves reflection with design and development, and in-engine playtesting. The loop has 4 core steps: A) Engaging in a chunk of design and development task until ready to test, B) Upon starting a playtest, the RDA tool prompts one to reflect on rationale and expectations of the test, C) Engaging in the playtest which the RDA tool records automatically, and D) Reflection on the playtest, prompted by the RDA tool when one stops the test.}
    \Description[A cyclic graph summarizing the RDA journaling process]{The graph contains 4 elements which link in order. 1. Design and development, Arrow text: The researcher reflects on the motivations or expectations behind the design test. 2. Pre-test comment, Arrow text: The researcher immerses themself in the in-engine test. 3. In-engine Test, Arrow text: Researcher ends test and journals reflection on the test experience. 4. Post-test comment, Arrow text: Gameplay Recording and Journal is saved.  Iteration continues.}
  \label{fig:RDA_Journalling_Loop}
  \end{teaserfigure}

\maketitle

\section{Introduction}



Digital games research is a broad and ever-growing field spanning multiple research domains. Building games as part of these academic endeavours is a common practice, but it currently faces two major limitations. Firstly, studies tend to overshadow the games and game design process itself, for example, a recent overview of academic games showed that games tend to serve as a means of data collection, test-beds for experiments, case studies, etc~\cite{chen_exploring_2025}. Secondly, HCI games studies tend to be overwhelmingly player-centric in nature, with design/designer knowledge largely missing~\cite{denisova_whatever_2021}, even though design research has established that there are unique insights to be gained through design practice~\cite{godin_aspects_2014, hook_games_2017}. As an example of knowledge gained through game design practice, Khaled and Barr~\cite{khaled_generative_2023} studied Barr's game design process and presented four theoretical ideas that helped capture Barr's personal game design approach and provide materialized evidence for known design dynamics.  Unfortunately, this kind of "designerly" game research appears under-represented and lacking in tools and processes.

In order to address these gaps, there is a need to support game-making efforts for studying various facets of design knowledge, namely, designerly knowledge contained in people (\textit{design epistemology}), practices (\textit{design praxiology}) and artifacts (\textit{design phenomenology})~\cite{cross_design_2007, kultima_game_2018}. Research-through-Design (RtD)~\cite{hook_games_2017, bardzell_documenting_2016, zimmerman_research_2007, gaver_what_2012, basballe_dynamics_2012, godin_aspects_2014} methodologies that aim to create knowledge through active design are well suited to address the aforementioned challenges. However, current methods and tools are not equipped to address immediate information capture, non-intrusive reflection journaling, granular design rationale capture~\cite{dalsgaard_reflective_2012}. 

To tackle these challenges, this paper proposes and evaluates Reflection at Design Actualization~(RDA), a novel tool for game design journaling, reflection, and data capture, implemented as a game editor extension for the cross-platform game engines Unity and Godot. RDA encourages moments of design reflection inside the game engine at points of designer playtests, which tend to happen frequently during day-to-day development. We argue that these playtests can be viewed as instances of design actualization, that is, moments when design decisions transition from concepts to tangible entities, and thus present a concentrated source for reflections.  

To summarize our contribution, the proposed open source RDA tool allows 
game designers and researchers to:

    \begin{itemize}
        \item Facilitate the collection of reflections with minimal intrusion to the design process itself.
        \item Record micro design reflections that capture rich tacit knowledge in a structured fashion.
        \item Track the holistic evolution of the project through automatic video recordings.
        \item Compile a dataset that is easily digestible and analyzable for the creation of design knowledge.
    \end{itemize}

We approach the development, analysis, and evaluation of RDA aligning with the processes of autobiographical design, which has been a prominent method to study design artefacts within HCI~\cite{luft_boards_2023, cao_dreamvr_2023, hammad_homecoming_2021, turchet_smart_2018, nunez-pacheco_sharing_2023, kaltenhauser_playing_2024}. Three designer-researchers engage with RDA in their artifact-based studies, with one project running for four months and two running for two months. 

As auxiliary contribution, we present findings regarding the designers' self-use experience with the RDA through three themes, namely, designer-routine compromise, designer-researcher persona consolidation, and the mirror effect of RDA. Additionally, we illustrate the designer experience through brief case studies and discuss current challenges. 

The ready-to-use package for Unity and Godot, along with the Python compilation script, is provided as supplementary material for those who would like to employ, extend or modify RDA in their own projects and processes.

\section{Background and Related Work}

\subsection{HCI, Design Research and Game Design}

HCI research is broad and multifaceted, though at its core, the goal remains to further the understanding between humans and technology. Thus, HCI studies often involve studying user experiences~\cite{cormier_this_2025, mirhadi_playing_2024, vakeva_dont_2025}, but this is only one side of the coin. It is equally important to understand how systems are designed, designer experiences, and understanding the artifacts themselves. This has been a space that design researchers have been actively exploring, with Cross~\cite{cross_design_2007} framing the study of designerly ways of knowing, design practices and artifacts as design epistemology, design praxiology and design phenomenology, respectively. The increasing prominence of autobiographical design methods in HCI research~\cite{kaltenhauser_playing_2024, nunez-pacheco_sharing_2023, cao_dreamvr_2023, cheng_reality_2023} may act as evidence of a symbiotic relationship between the two domains. 

Autobiographical design has its methodological limitations, the greatest being that it requires researchers to be genuine users of the system they are designing~\cite{neustaedter_autobiographical_2012}. This makes autobiographical design challenging to apply to domains like games, where the game designer is not truly a `user' of the game that they are crafting. Research through Design (RtD) is another umbrella term that captures knowledge generated by studying the materials, processes, and artifacts of design~\cite{frayling_research_1994} without the explicit need of a user. Godin and Zahedi~\cite{godin_aspects_2014} add that RtD involves crafting artifacts that cannot be fully described and promote a dialogue between the designer and the situation, leading to learning experiences. They also emphasise that the goal of RtD is not the artifact itself, but rather knowledge and understanding gained through the act of design. 

Recently, Bates and Kirman~\cite{bates_making_2025} have made a similar argument for Research through Game Design, which involves engaging in the act of making games to learn and generate knowledge. They add that games need not necessarily be built to be measured, sold, played, or even finished (the opposite of which is the norm in academic games~\cite{chen_exploring_2025}) to contribute novel insights. They state that knowledge generation happens during the very process of engaging with game design. Academia offers a unique avenue to build and learn from games which might otherwise never be built or studied. Khaled~\cite{khaled_questions_2018} echoes this, stating how both serious games and entertainment games present limitations to designers' freedom to explore less conventional practices. We don't want to frame that there is little value in studying the design of serious or entertainment games, but rather that there needs to be equal emphasis and support for experimental game design~\cite{khaled_questions_2018}.

\subsection{Game Making in Games Research}

There has been a growing interest in understanding designer perspectives within games HCI. Denisova et al.~\cite{denisova_whatever_2021} explored how game designers create and curate emotional experiences in emotionally impactful games through an interview study, highlighting that the designer perspective is largely overlooked in games HCI. Hicks et al.~\cite{hicks_juicy_2024} similarly interviewed game designers to understand how they conceptualize juicy audio, i.e. innately satisfying audio, in games. The designer perspective was crucial here since `juiciness' is a relatively fuzzy term, and the craft surrounding it largely involves tacit knowledge that is hard to describe. Bhatnagar et al.~\cite{bhatnagar_beyond_2025, bhatnagar_understanding_2024} explored another fuzzy game design term called `game feel'. Although their study revolved around understanding game-playing experience, they chose to recruit participants with game design experience and supplemented the data with personal design discussions because there again, the tacit knowledge proved crucial to the exploration of the concept. 

Game making is also a common activity within games HCI. As an example, Gómez-Maureira et al.~\cite{gomez-maureira_level_2021} built an exploration game with four versions to explore how level design patterns relate to curiosity-driven exploration. The game design may need not be be high-fidelity. For example, Hammad et al.~\cite{hammad_homecoming_2021} explored the experience of returning to first-person games after a long pause. As part of their process, they built a small level by modding the popular game The Witcher 3. In Virtual Reality studies, games are often developed to, for example, explore novel types of exergaming  ~\cite{kontio_i_2023} and novel social interactions like pair-dancing~\cite{laattala_duet_2024}, or to translate familiar mechanics from conventional games into VR~\cite{kirjonen_souls-vr_2025}. 

Though game-making is common and there is a desire to better understand designer perspectives, games are seldom made as a reflective design practice. We might better understand the practices of games on a cultural or industrial level~\cite{kultima_game_2018}, but there is a lot of unexplored value in understanding personal design experiences.

\subsection{Documenting Design}

Documenting design knowledge is a well-recognised challenge, and various attempts have been made to tackle it. Bracewell and Wallace~\cite{bracewell_tool_nodate, bracewell_capturing_2009} built the Design Rationale editor, a software to capture design rationale. Their preliminary findings were positive but as the evaluation took place in the specific industrial context of aerospace, the implications to game design are unclear. Dalsgaard and Halskov~\cite{bardzell_documenting_2016} developed the Project Reflection Tool, a generalized framework to document the research through design process, which acts as a close parallel to our work. Technically speaking, it differs in being a more general web-based system instead of a dedicated tool integrated in a game editor. In terms of practice, the biggest difference is that their system captured bigger design decisions around design activities called `events' that span a day's work, whereas we capture reflections before and after each time the designer starts the game in the game editor. They state that in their project, their main goal was not to present the system, but rather to share their experiences using their tool as a catalyst for design knowledge in a variety of use cases. Our approach aligns with this mindset as well, where the designer's experiences with the tool are more important than the raw efficiency of the tool. When reflecting on the challenges of their tool, Dalsgaard and Halskov~\cite{bardzell_documenting_2016} note that their tool would benefit \textit{“primarily with regard to the immediate capture of information throughout the design process”}, which is a major aspect explored in our work.  

This challenge of documentation has been acknowledged in the game design as well. Khaled et al.~\cite{khaled_documenting_2018} attempted to study personal and nuanced experiences of game design as they may exist "in the wild". Their approach involved studying the version control commit logs of the game design to explore the design journey and understand motivations for various design actions. This process was eventually concretized as the Method of Design Materialization (MDM)~\cite{khaled_generative_2023}. MDM utilizes the version control repository as a design database, which is built organically as the designer engages with it rhythmically and chronologically during the game design process~\cite{khaled_method_2023}. Khaled and Barr~\cite{khaled_generative_2023} utilized this method to study Barr's design process in the game \textit{It is as if you were making love}\footnote{https://pippinbarr.com/itisasifyouweremakinglove/}, and the process was able to yield insightful details regarding Barr's approach to game design. To further assist in this process, Llagostera et al. developed a desktop app tool called Ponte~\cite{granzotto_llagostera_tracing_2025} to act as a bridge between the design materials present in version control systems and qualitative data analysis software. The tool allows for chronological analysis for various version control metadata, which otherwise remains invisible, such as file addition/deletions and line modifications.  This, along with tagging and filtering, lets researchers study the design process with additional context and flexibility. Ponte further allows this data to be exported to qualitative data analysis software for further analysis.

\paragraph{Situating RDA Within the Literature }

In summary, our RDA approach extends the work of Ponte~\cite{granzotto_llagostera_tracing_2025} and Khaled et al.~\cite{khaled_documenting_2018, khaled_generative_2023} by providing a new data source (explicit granular designer reflections and video capture) that could augment the version control data that they used. We also extend Dalsgaard and Halskov~\cite{dalsgaard_reflective_2012} by integrating our journaling tool in a game editor and automatically prompting the designer for reflections whenever they test the game.

\section{Reflection at Design Actualization}
\label{sec:4.0}

This section describes the proposed RDA tool and process in detail. We begin by discussing the origins and requirements of RDA, followed by the specifics of the RDA process as well as the RDA tool implementation details. Finally, we discuss the dual nature of RDA as a tool and process. The methodology we used to develop and evaluate RDA is described later in Section~\ref{sec:methodology}.

\subsection{Project Origin}
\label{sec:4.1}

Following recommendations by Desjardins and Ball~\cite{desjardins_revealing_2018} on autobiographical design, we briefly discuss the origin of the project in order to contextualize project needs and add to the credibility of the findings. R1's primary research focus has been a phenomenon within game design known as \textit{game feel} that can be understood as the moment-to-moment sensational experience of playing a digital game ~\cite{swink_game_2009, pichlmair_designing_2022, bhatnagar_beyond_2025, game_makers_toolkit_secrets_2015}. This phenomenon captures a unique experiential dimension for both players and designers, and is very much centred around the micro-design decisions that create a noticeable impact on player experience~\cite{nicolae_berbece_game_2015, the_game_overanalyser_art_2019, pichlmair_designing_2022}. 

Despite the above, most studies on game feel have been retrospective in nature and focused primarily on the player and not the designer. This gap, in addition to R1's interest in research through game design methodologies, led to a plan to study game feel through a first-person game development project. During a supervision meeting between R1 and their supervisor, they agreed that no existing method fit the requirements of this kind of data capture, namely tracking micro-design reflections and visual tracking of project evolution, forming the initial requirements of RDA. It was thus decided to build a simple tool that can record the in-engine playtests along with a pre-test and post-test reflection.

From this point on the tool and its requirements grew progressively, from its inception on 02.04.25\footnote{Note: all dates mentioned in the paper are in dd.mm.yy format} until the writing of this article on 01.09.25. We discuss this evolution and the final tool details in the following sections.

\subsection{Needs/Requirements}
\label{sec:4.2_requirements}

The following are the goals of RDA, along with the context in which they were conceived: 

\begin{enumerate}
    \item \textit{Capturing micro-design reflections}: This was one of the foundational requirements before any development of the project on 02.04.25.  The tool need not exclusively capture micro-design reflections, but it should have the capability to.
    \item  \textit{Automatic video recordings}: This requirement can be seen as a dual pair of the first requirement, both being envisioned in parallel (on 02.04.25) and feeding into each other. Since the micro-design reflections can be quite specific, they may prove challenging to analyze with their chronological design context. These video references would also act as growth trackers, helping researchers further understand design(er) behaviours and influences of decisions. 
    \item  \textit{Minimizing intrusion to the design process}: This requirement developed progressively during early tool usage when the lack of quality-of-life features in the prototype started adding up to a noticeable amount of friction (between 30.04.25 - 17.06.25). Though zero friction is not possible, or even desirable, since some amount of friction helps to switch to a critical thinking mode, the experience needed to be smoothed out enough to avoid being felt as intrusive.
    \item  \textit{Processing raw data to be digestible and analyzable}: This was the last major requirement, which was developed (approx. on 30.06.25) once the tool use had already rendered some volume of raw data. Although rich, it proved overwhelming to wrangle, so some means to make the data manageable for qualitative studies had to be developed.  
\end{enumerate}

\subsection{The RDA Process}
\label{sec:RDA_method}

\subsubsection{Design Actualization}

In the context of design, to actualize is to make something that is possible into something real. Design Actualization are moments in the design and development process when abstract design elements, such as tacit knowledge, ideas, mental concepts, experimental or exploratory thoughts, materialize through a design decision. Prototypes and sketches are a very common design artifacts that can be seen as actualization of design ideas.  For example, in the Drift Table project~\cite{gaver_drift_2004}, various 2D and 3D renders were used explore ludic engagement, each acting as an instance of design actualization. While a prototype gives the notion of some degree of completeness, design is actualized in smaller steps all the way to the completion of a prototype, or in the journey to a failed prototype. These smaller actualization moments are particularly prominent in design fields with faster iteration cycles, such as games, creative coding and new media, where small choices are made and tested dozens to hundreds of times before the completion of a prototype. In digital game development, this design actualization tends to be the moment inside the game engine when the game is run (by a designer/developer) to test a feature, effect, level, feel or anything else. Designers/developers tend to innately adopt this as a ritual of game development, each test serving as a reflection moment from where the future direction is calibrated. It is these reflections that we argue are rich, untapped sources of design knowledge that are rarely documented and studied. For example, in a game with a controllable character movement, multiple acceleration values might be tested and adjusted to achieve a desired effect, each test being an instance of design actualization. Within the temporal vocabulary developed by Oogjes and Desjardins ~\cite{oogjes_temporal_2024} , moments of design actualization can be seen as "Encounters" between the designer, the concept and its tangible representation. The way these moments repeat thoughout the design process make them suitable for reflective practice. The term actualization is synonymous with the term materialization, as used in the Method for Design Materialization~\cite{khaled_generative_2023, khaled_documenting_2018, granzotto_llagostera_tracing_2025}. Although there are overlaps, we see materialization as an umbrella term to indicate anytime a design activity leaves a trace~\cite{khaled_generative_2023, granzotto_llagostera_tracing_2025}, while we use actualization as a specific materialization moments that can be experienced. For example, code refactors or designer ponderings/journaling notes are instances of materialization but not necessarily design actualization since a design decision is not being experienced.

\subsubsection{The Reflection at Design Actualization Process}

The RDA process can be employed though the following steps:

\begin{enumerate}
    \item  Establish a research goal or direction for the project, as one would in any research-through-design undertaking~\cite{granzotto_llagostera_tracing_2025, gaver_what_2012, khaled_generative_2023, khaled_documenting_2018, gaver_drift_2004}.
    \item Identify the relevant point(s) of design actualization, in games this is generally the in-engine test moment.
    \item Engage in the RDA iteration, visualized in Figure~\ref{fig:RDA_Journalling_Loop}, as follows:
            \begin{enumerate}[label=(\Alph*)]
                \item Perform design and development tasks until something can be tested or experienced.
                \item Right before interacting with the designed artifact, reflect on the motivations and expectations of the design decisions and/or what you are hoping they accomplish in the test. Our RDA tool prompts this automatically.
                \item Immerse yourself and/or participants with the artifact. This interaction should be captured or recorded in some fashion (the RDA tool automates this recording step). Note that some design decisions may happen during this step itself. They can be captured in the recording, as a voice memo, or as a post-test comment as explained in the next step.
                \item Right after this test, reflect again on the experience, how it compares to your thoughts on the pre-test reflection, any changes that happened during the test and any future design inspirations or thoughts. Prompting of this comment is again enabled by the RDA tool.
            \end{enumerate}
            \item Repeat step 3 until project completion or when you are satisfied with the data. We also recommend maintaining a design journal throughout the process to capture any relevant thoughts that may happen outside of this loop.
\end{enumerate}

This process is quite fundamental, but at the same time it is a delicate balancing act between the researcher and designer persona within the same actor. Moments of design actualization are when the designer persona tends to be most prominently in control, so to spotlight the critical researcher can potentially hamper the design process itself. There is no clean solution, but as discussed in detail in the results (Section \ref{sec:results}), this friction of gear-shifting between the designer and researcher persona can be minimised with appropriate tool design and practice on the part of the actor.   

Note that we do not claim that moments of design actualization are the only valuable sources of gathering design knowledge, nor that it can be applied in every design study and scenario but that it is one option that should be  considered. We also want to avoid methodological prescriptivism (there is no "correct" or "incorrect" RDA approach), the aforementioned steps are \textit{our} conceptualization of the RDA process and we encourage others to adapt and experiment in a well-reasoned fashion.

To form insights from this process, one should pair RDA with some form of analysis techniques such as Thematic Analysis~\cite{braun_toward_2023, braun_using_2006, byrne_worked_2022}, Grounded Theory~\cite{charmaz_constructing_2012, salisbury_grounded_2016, cole_more_2022} etc. This is in line with current developments in RtD, for example, the Method of Design Materialization~\cite{khaled_documenting_2018, khaled_generative_2023, granzotto_llagostera_tracing_2025} also utilizes journaling and reflection, with Grounded Theory as their recommended analysis method.

\subsubsection{RDA Tool and Process Duality}

Throughout the paper, we discuss RDA as both a process and a tool. RDA process is the notion of identifying design actualization moments, engaging in reflection around the moments and recording interactions during them. The RDA tool is an interface that facilitates this process, here implemented as a game engine extension. For the authors, RDA was experienced as an amalgam of both, since the game engine extension was the only way though which the process was interfaced. Through autobiographical design, we present not just a tool that fulfils the goals of RDA, but also the process and challenges involved. This additionally helps us highlight how interactions with the tool directly influence how the researchers experience the process. We use the term RDA throughout to discuss this dual natured experience and specify tool or process when discussing one side explicitly.

\begin{figure}
    \centering
    \includegraphics[width = 0.95 \linewidth]{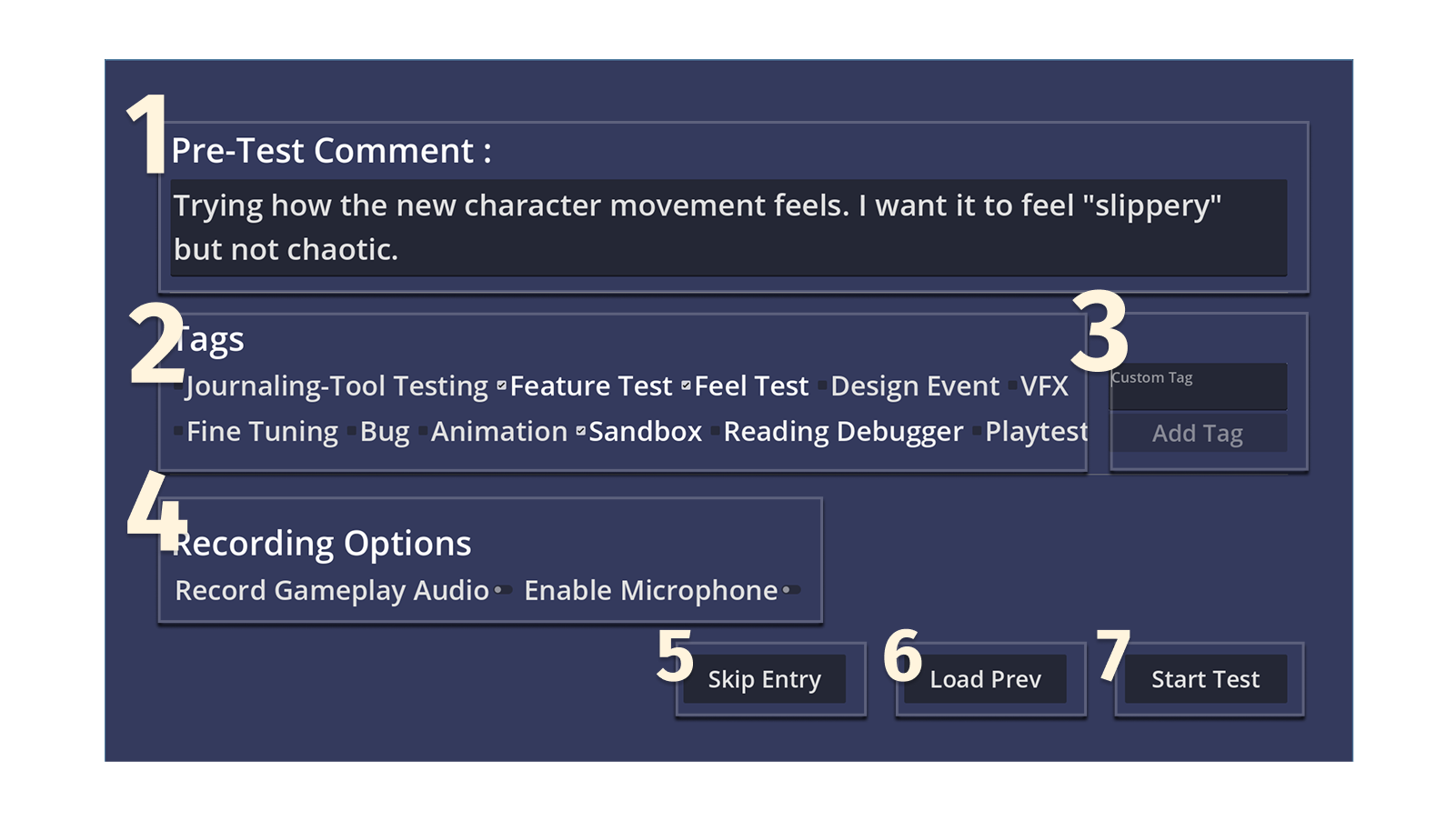}
    \caption{Breakdown of the distinct components of the RDA tool UI. 1. Reflection input field, 2: Tag selection check boxes, 3: New tag creation input field and button, 4: Video recording options, 5: Skip journaling entry button, 6. Load previous entry button, 7. Start test button.}
    \Description[Layout of RDA Tool  UI layout]{The visual layout of the is roughly in thirds vertically. The first third if the screen contains the reflection input (1). The second third contains the tag checkboxes and tag creation area (2 and 3). The final third contains the recording options, and buttons (4, 5, 6 and 7).}
    \label{fig:tool_breakdown}
\end{figure}

\subsection{RDA Journaling Tool Details}
\label{sec:4.4}


The technical implementation combines the following components: 

\begin{itemize}
    \item \textit{Game Engine of Choice}: The tool is centred around the use of a game engine where the design, development and testing of the game takes place.  In our current prototype, we have built and tested the tool on \textit{Godot}\footnote{https://godotengine.org} and \textit{Unity}\footnote{https://unity.com/products/unity-engine} , which are two of the most popular game engines in hobbyist, solo and indie development spheres\footnote{https://itch.io/game-development/engines/most-projects}\footnote{ https://bsky.app/profile/gamemakerstoolkit.com/post/3lvkrymak5s26}. In theory, the tool should be adaptable to any game engine with in-game UI, but the communication between OBS (see below) and the game engine may prove to be technically challenging. Said communication was simplified in Unity and Godot thanks to community-built extensions to enable and simplify Websocket communication.
    \item \textit{OBS}: The Open Broadcasting Software Studio (OBS)\footnote{https://obsproject.com} is an open source video streaming and recording tool. We use it for the automatic recording part of the tool since it comes with in-built WebSocket Plugin, which is a communication protocol that allows to easily exchange between OBS and the game engine. 
    \item \textit{Python}: We use a simple Python script to help convert the raw data collected using the tool into clean compilations that are easier to analyze.
\end{itemize}

\begin{figure}
    \centering
    \includegraphics[width = \linewidth]{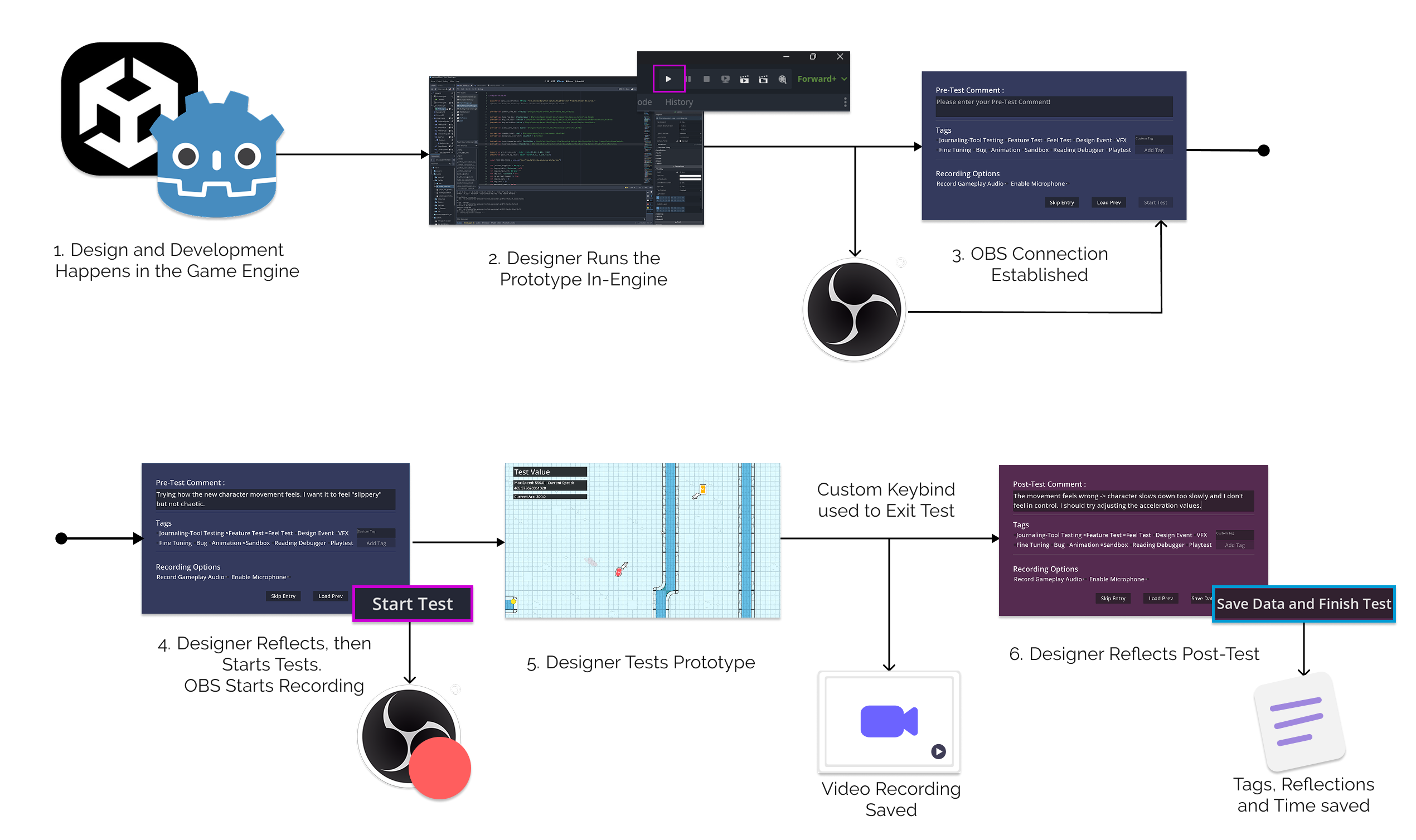}
    \caption{The flow of a single playtest with RDA journaling.}
    \Description[A 6 step linear flow]{6 image elements which connect linearly. 1. The logos of Unity and Godot, stating that design and development happens in the game engine. 2. A screen shot of the Godot engine, highlighting that the developer is pressing "Play" after finishing a design task. 3. A screenshot of the Journaling UI and Showing that the OBS connection has been established. 4. Designer journals reflection in the journaling UI and presses the Start Test buton. OBS starts recording. 5. A screenshot of the game being playing inside the game engine. 6. The designer enters a custom keybind which stops the recording and saves the video. A screenshot of the post-test reflection UI is shown, highlighting that the designer presses "Save Data and Finish Test" button which saves the reflection as a json file. }
    \label{fig:tool_flow}
\end{figure}

With the components in place, we now discuss how a sample iteration of the RDA plays out technically. A visual summary of the process can be seen in Figure~\ref{fig:tool_flow} . The process has only been conducted on the Windows operating systems. The steps should be transferable to other operating systems, but validation is beyond the scope of the current study. An initial setup is required for both OBS Studio and the game plugins, which are skipped from the current description in favour of brevity, but those details can be found in the supplemental.

At the start of a dev cycle, OBS Studio is launched and passively exists as a background process, handling recording when needed. OBS can be calibrated to record just the gameplay, the entire engine or even other screens/camera feed. Inside the game engine, the in-game User Interface (UI) is used for reflection input. Note that this UI does not refer to the UI of the game engine, but rather a UI that runs as part of the game prototype itself. This decision was made for a few of reasons: 1) it reduces friction by keeping the entire implementation within the game engine, 2) the OBS communication and reflection management runs as a game script and thus it needs the game to be running to operate, 3) this keeps the implementation engine-agnostic and easily customizable, and 4) it's the easiest way to ensure the reflections are presented before and after the test. We acknowledge different implementation possibilities, but this method proved the most consistent for us and we invite further experimentation and improvements.

Once the designer is done with a design or development task, they usually test out their work by running the game in-engine (see Figure~\ref{fig:tool_flow}, step 2). As soon as this test is started, the OBS connection is established (see Figure~\ref{fig:tool_flow}, step 3). Using OBS can be avoided altogether in Unity using the \textit{Unity Recorder Plugin} , but it comes with its own limitations. R1 used OBS since they used Godot for development, R2 started with Unity Recorder but switched to OBS, while R3 did their entire journaling in Unity Recorder. 

With the game prototype launched, the designer may do a pre-test reflection. The nature of this reflection depends on the goals of the designer and project, they might reflect on a specific theme or aspect of design, how the current decision aligns with their design goals, or it may be more exploratory in nature. The reflection itself is just entered in an input-field. Tags may also be created and assigned at this step. There is an option to load the previous reflection since which can help reduce friction when the changes or tests are similar. This can happen often in the current context of smaller design decisions. The designer may also choose to skip the journaling step at this point if they feel that the current test step does not add value, which can be the case when fixing bugs, for example. This is a matter of designer choice and discipline, sometimes a reflection's value is only seen in retrospect, but depending on the designer's work style, recording everything may bloat the data. If the designer chooses to reflect and start the test, the RDA tool sends OBS a Websocket message to start recording (see Figure~\ref{fig:tool_flow}, step 4). Once the recording is confirmed, the game enters a testing session (see Figure~\ref{fig:tool_flow}, step 5). The OBS icon in the taskbar also changes to indicate that a recording is active and having OBS open on a second monitor is even better to check if the recording happens correctly. 

When the designer is done with the test, they have to enter a custom key-bind, for example, R1 used \textit{Ctrl + L} in their project, to trigger the end of test and start the post-test reflection (see Figure~\ref{fig:tool_flow}, step 6). This is done because the \lq end of test\rq~is non-trivial to detect in most game-engines without stopping the execution of game scripts, and as discussed earlier, the OBS communication and reflection systems are dependent on that. This requirement is a deviation from the general designer muscle memory, which involves ending the test using a stop button in the engine UI. Note that if the designer forgets to hit the keybind, a journal log will not be created, and OBS will have to be stopped manually. We chose this as a tradeoff compared to the alternative of collecting reflections out-of-engine as that would increase the number of applications to manage and also decouple the reflection from when the in-game test starts. In the post-rest reflection, the designer may compare their experience with the pre-test reflection, modify tags and then end the test.

Once the reflection is complete, a JSON file containing the tags, pre-test comment, post-test comment and the respective date and time is created. An MKV video (MP4 in case of Unity Recorder Usage) of the test is also saved. These files are stored in a directory named by dates and then subdirectories named with their respective run number. A new folder is automatically made for each day and then the respective run. These steps are repeated as long as the researcher desires.

\begin{figure}
    \centering
    \includegraphics[width = \linewidth]{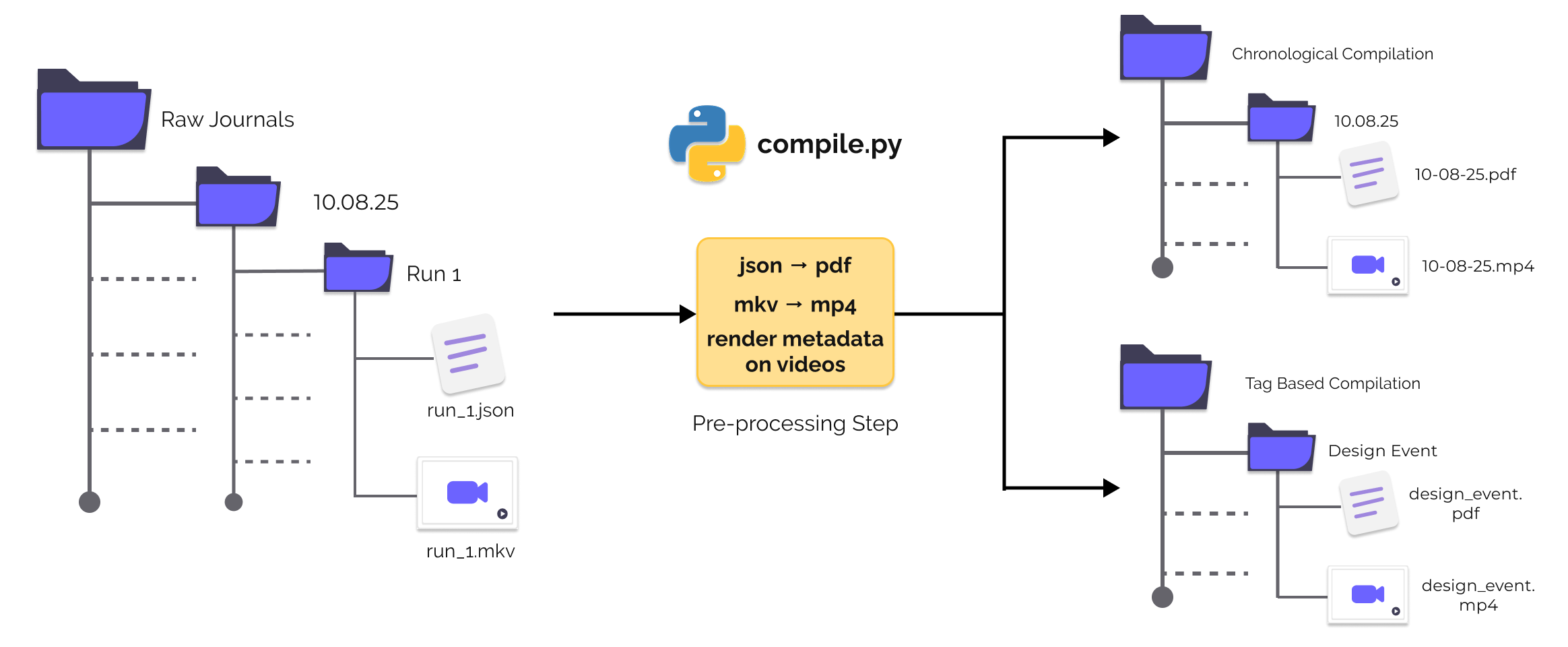}
    \caption{The compilation process flattens the collected raw data (playtest videos and notes) into a more manageable form. In the chronological compilation, every day's logged reflection and video gets cleaned and compiled into a single pdf file and mp4 video respectively. Similarly, in the tag based compilation, every entry containing the tag gets compiled into one pdf file and video.}
    \Description[A linear flow illustrating the compilation process.]{The figure has 3 distinct sections horizontally. Section 1 shows the file structure of the raw journals. There is a 3-level hierarchy: Raw journals folder containing folders of multiples dates. Each date folder containing folders for multiple runs. Each run folder containing a single json file and mkv video. Section 2 shows the python script doing a pre-processing step, then doing the compilation process. In the pre-processing, it converts jsons to pdfs, mkvs to mp4s and renders metadata on the videos. Section 3 shows the the 2 results of the compilation. Each one has a 2 level folder hierarchy compared to the 2 level of the raw data.}
    \label{fig:tool_compilation}
\end{figure}

\subsection{Compilation Script Details}

\begin{figure}
    \centering
    \includegraphics[width = \linewidth]{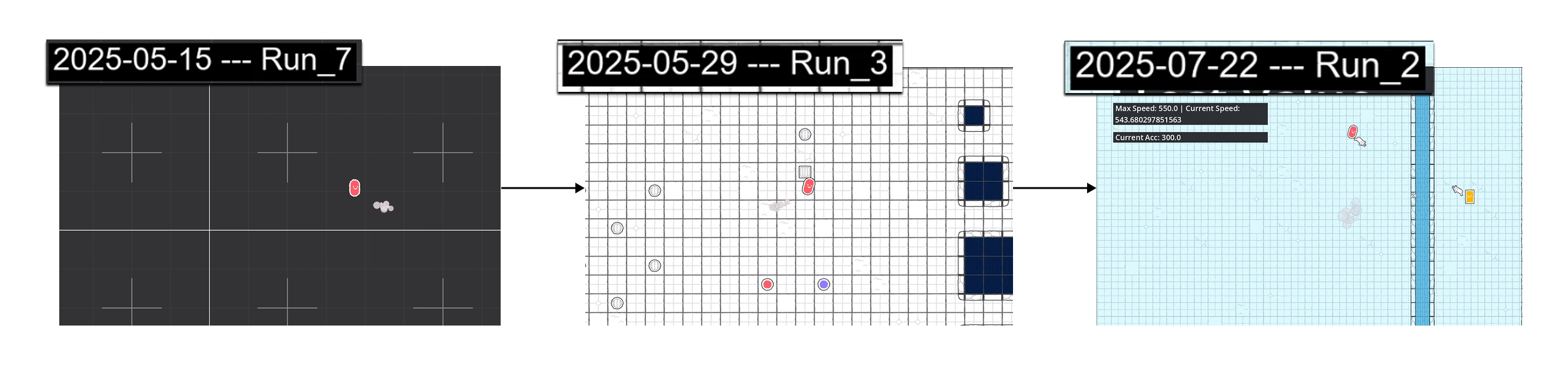}
    \caption{An example of how the RDA tool helps capture project evolution illustrated through video captures from R1's project. Note the metadata that is rendered on the video as part of the compilation process.}
    \Description[3 video frames from R1's game project.]{The figure shows 3 video frames, each almost a couple of weeks apart. The first one only has the player character, the second one has some level elements. The third one a bigger level, more fleshed out level and an enemy. The date and run numbered is rendered in each video}
    \label{fig: project_evolution}
\end{figure}

The raw data generated by the RDA tool above can be too dispersed to analyze efficiently. A Python script was developed to process the data into a more digestible form (see Figure~\ref{fig:tool_compilation}). The script first does a pre-processing step, converting the raw json into a readable pdf. Additionally, the date and run number are rendered on the video itself (see Figure~\ref{fig: project_evolution}), and it is converted to an mp4 format.

The script then generates per-day compilations, concatenating all the reflections and videos into a single pdf and mp4, respectively. Additionally, target tags can be specified to compile journal entry tags used throughout the journaling process. These tag-compilations can be done using a logical AND operation (compiling entries where all the target tags are present) or a logical OR operation  (compiling entries where any of the target tags are present). To help make an informed decision on which tag combinations to compile, a conditional probability heatmap is generated to give an idea of what tags tend to be used together often. 




\section{Methodology}\label{sec:methodology}

This section describes the methodology we used to develop and evaluate RDA. For details on the RDA method itself, see Section~\ref{sec:RDA_method}. We engaged in Autobiographical Design (AD) during the process of building and evaluating both the RDA tool and the associated process. The data collected through the autobiographical design process was analyzed through Reflexive Thematic Analysis (RTA). Below, we elaborate on our rationale and approach towards Autobiographical Design, the researchers' positioning, and the RTA details.

\subsection{Autobiographical Design}
\label{sec:3.1}

AD is a first-person study approach often used in Design and HCI research for artifact-focused studies. We adopt AD for this study based on three key rationale. Firstly, AD distinguishes itself from other first-person approaches, such as autoethnographies, by being centred around the design of one or more artifacts, and the researcher taking the role of the designer and user of the system~\cite{kaltenhauser_playing_2024, neustaedter_autobiographical_2012}. Neustaedter and Sengers~\cite{neustaedter_autobiographical_2012} frame AD as \textit{"design research drawing on extensive, genuine usage by those creating or building the system"}, where genuine refers to the design guided by the needs of the researcher rather than an external user. Throughout the development of RDA, the three researchers have been true users of the system, employing RDA in their artifact-based research processes. Secondly, AD allows for the weaving of personal narratives into the design and evaluation of technology \cite{kaltenhauser_playing_2024}. We highlight our personal and cultural positioning as (game) designers who are also acting as researchers engaging in varied research processes that involve the creation of a game (or interactive) artifact. Finally, our work fits key practice characteristics of AD set out by Neustaedter and Sengers~\cite{neustaedter_autobiographical_2012}, namely, fast tinkering, operating with real systems and leveraging long-term usage.


We briefly note again that although Autobiographical Design and Research through Design (RtD) share methodological features, the above rationale acts as a key motivator for us framing our study as AD over RtD. We highlight that in RtD the researchers need not be a true user of the system or that the concept of a user need not even apply. Additionally where AD approaches are driven by designer-researcher needs and requirements, RtD approaches tend to lean more on evolving Manifestos and Artistic/Design Statements which tend to capture the direction of the process over end goals~\cite{granzotto_llagostera_tracing_2025, gaver_what_2012, khaled_generative_2023, khaled_documenting_2018}.



\subsection{Author Backgrounds and Positioning}
\label{sec:3.2}

Autobiographical Design is centred on the researcher's personal experiences and therefore we consider it important to briefly understand the researchers as actors and contextualize the subjective insights in the following sections. The first three authors engaged in the use and evaluation of RDA. The first author acted as the primary developer and designer of the RDA tool and the primary investigator of this project. The pronoun `we' is used throughout the project to reflect the shared perspective of every author. When the personal experience of a single author is addressed, a coded pronoun is used instead, for example, \textit{`R2 states that ...'}. Here, `R' stands for researcher, and the number is their author index. 

\paragraph{First Author}
The first author, referred to as R1 throughout, is a game designer, programmer and researcher. They are a part of a games research group, engaging in mostly qualitative research, and actively developing digital games as part of their studies and as a personal hobby. Their game development experience has been in both solo and team settings, but they lack industrial development experience. Their research goals aim to study the design processes of video games, through the process of building games, bringing emphasis on the designer. This project is a direct response to their aforementioned research goal.

\paragraph{Second Author}
The second author, referred to as R2 throughout, identifies as a new media artist and Extended Reality designer, with a focus on player experience and interaction design. Their research examines embodied perception in virtual environments, conceptualized through the technological loop between body, technology, and environment. They also investigate the performative recontextualization of sound in VR as a generative process rather than reproduction. R2’s work is situated at the intersection of artistic practice and research, centering on player’s embodied interaction, emotional experience, and feedback. This project directly contributes to these aims by offering a space to study how design choices shape the affective and experiential dimensions of play.

\paragraph{Third Author}
The third author, referred to as R3 throughout, identifies as a game designer, programmer and researcher. They are also a part of a games research group, where they focus on virtual reality and computer vision-based locomotion and interaction design. Their research aims to contribute to the creation of new locomotion and interaction techniques and study their applicability in exergames. Prior to their studies and research related to game design and development, they have experience working in the mobile games industry, where they successfully developed and shipped multiple products for the Android and iOS platforms. This project contributes to exploring the design and development process of a novel exergame locomotion and interaction technique in an academic setting.

\subsection{Procedure}
\label{sec:3.3}

Each researcher used the RDA tool independently in their respective projects (which have been detailed in Section \ref{sec:4.3_case_studies}), being critical throughout and reflecting on how RDA influences their workflow. The meta-reflections were chronologically journaled and maintained throughout the span of the project. Note that throughout this section (Section~\ref{sec:3.3}), the mention of reflections refer to the meta-reflections that the authors made regarding RDA and not the design reflections on their projects through the use of RDA. We use the term meta-reflections to clarify that distinction. The designers did not necessarily aim to finish said projects during the duration of the study, rather, the goal was to immerse themselves with RDA in order to evaluate its pros and cons. The authors worked in co-situated spaces and often discussed RDA and their projects with each other. 

R1 started designing and building the RDA tool on 02.04.25. R1 considered the tool usable on 30.04.25. On that day, R1 started actively developing their artifact-based project and actively using and improving the tool. R1 continued using the tool till 22.07.25 at which point they started working and researching on this paper. R2 and R3 started using the tool on 01.07.25 and continued using it till 28.08.25.

At the end of the meta-reflection collection process, R1 drafted survey questions to capture each participation researcher's overarching retrospective thoughts and experiences with the use of the RDA tool. This survey was answered by each author independently, including R1, who drafted and answered the questions before starting to analyse the data from the other designers. R1 then conducted Reflexive Thematic Analysis~\cite{braun_using_2006, byrne_worked_2022, braun_reflecting_2019} on both the meta-reflections and the survey responses to construct themes that captured the holistic designer experience.

\subsubsection{Reflexive Thematic Analysis}

We chose RTA, a qualitative data analysis method used for the development of themes or patterns that capture important aspects of the data~\cite{braun_using_2006, byrne_worked_2022}, to capture the lived experience of the designers. This method was chosen over other established Thematic Analysis (TA) methods, such as codebook TA and coding reliability TA which promote objective or structured coding, because reflexive TA provides interpretive power to the researcher and promotes contextual sense-making~\cite{braun_reflecting_2019}. This was considered crucial to generating meaning unified interpretive stories~\cite{braun_toward_2023} that capture the personal narratives of the designers. The TA was conducted solely by R1. Their theoretical stance is rooted in contextualism and constructivism. Through contextualism, they believe that knowledge generated can be best understood within the context it exists in, in our case, it is the designers' contexts. Through constructivism, they believe that knowledge does not exist latently in the data, but rather that meaning is constructed by the researcher and is shaped by their social and cultural environment.

The 6-step process outlined by Braun and Clarke~\cite{braun_using_2006} was followed to conduct our TA using Atlas.ti\footnote{https://atlasti.com}. The coding of the data was done inductively, with no codes being determined before analysis. Yet it should be reiterated that this process was still influenced by R1's background as a games researcher and game developer. Codes were constructed at both a semantic/explicit and latent/interpretive level. Examples of semantic codes include \textit{"tool making designer self-conscious"} which was explicitly stated by the designer. Meanwhile, examples of latent codes include \textit{"requiring designer to change habit"}, where this fact was not explicitly stated but could be interpreted from the context of the data. After initial coding, codes were then grouped, merged and relationships established. Atlas.ti's network feature was then used to arrange codes and form themes. Topic summaries were avoided~\cite{braun_one_2021, byrne_worked_2022} in favor of themes that capture the overall designer experience, which better encapsulate why various experiences originated and how they influenced the designers. Two iterations were conducted before the 3 themes were finalized, named and reported.

\section{Results}
\label{sec:results}

The results are presented in 3 perspectives. Firstly, we present each researcher's project and processes as case studies to contextualize out findings. Following that, we share the key themes that were constructed to capture the lived experience of the designers while engaging with RDA. Finally, we compare the feedback on RDA use against our design goals, highlighting its strengths and challenges.

\subsection{Case Studies}
\label{sec:4.3_case_studies}

\subsubsection{A Design Exploration of Game Feel}

R1's project involved building a simple 2D action game. Their design goal was to explore and understand game feel from a designer perspective, though studying their own micro-design decisions. They started building the RDA tool before the project itself, and the tool's initial design was driven by their general needs to improve the tool's user experience (like the tagging feature and automatic directory management) to alleviate major moments of friction. As the tool use became more intuitive, they started developing new rituals around the tool, like starting each day with a test with no changes and reflecting on how they felt about the prototype. R1 used rationale based questioning during their reflections, attempting to explicate their "gut-feeling" decisions. They also incorporated temporal vocabulary~\cite{oogjes_temporal_2024} as a part of their design and analysis process. Design insights also started forming as a part of the reflection process itself, and R1 would usually encounter these "critical tests" which would lead to design knowledge breakthroughs during post-test reflections.

One of R1's biggest frustrations came from not being able to work on the game because of the time spent building, researching and developing RDA. They enjoyed engaging with RDA to study their process and want to engage with it more. At the end of the autobiographical design, they are optimistic about continued RDA use in academic endeavours, but they would avoid using it in non-academic endeavours. They believe critical reflections and intuition-driven design both have their own time and place. Given their extended use RDA academically, they would like to design intuitively within their personal projects. 

R1 recorded 186 logs over 23 sessions. The total data size was 7.5 GBs. This count ignores about 10 sessions which where spent during the initial tool development where no analysable data (in terms of RDA reflections) was being collected.

    \begin{figure}
    \centering
    \includegraphics[width = \linewidth]{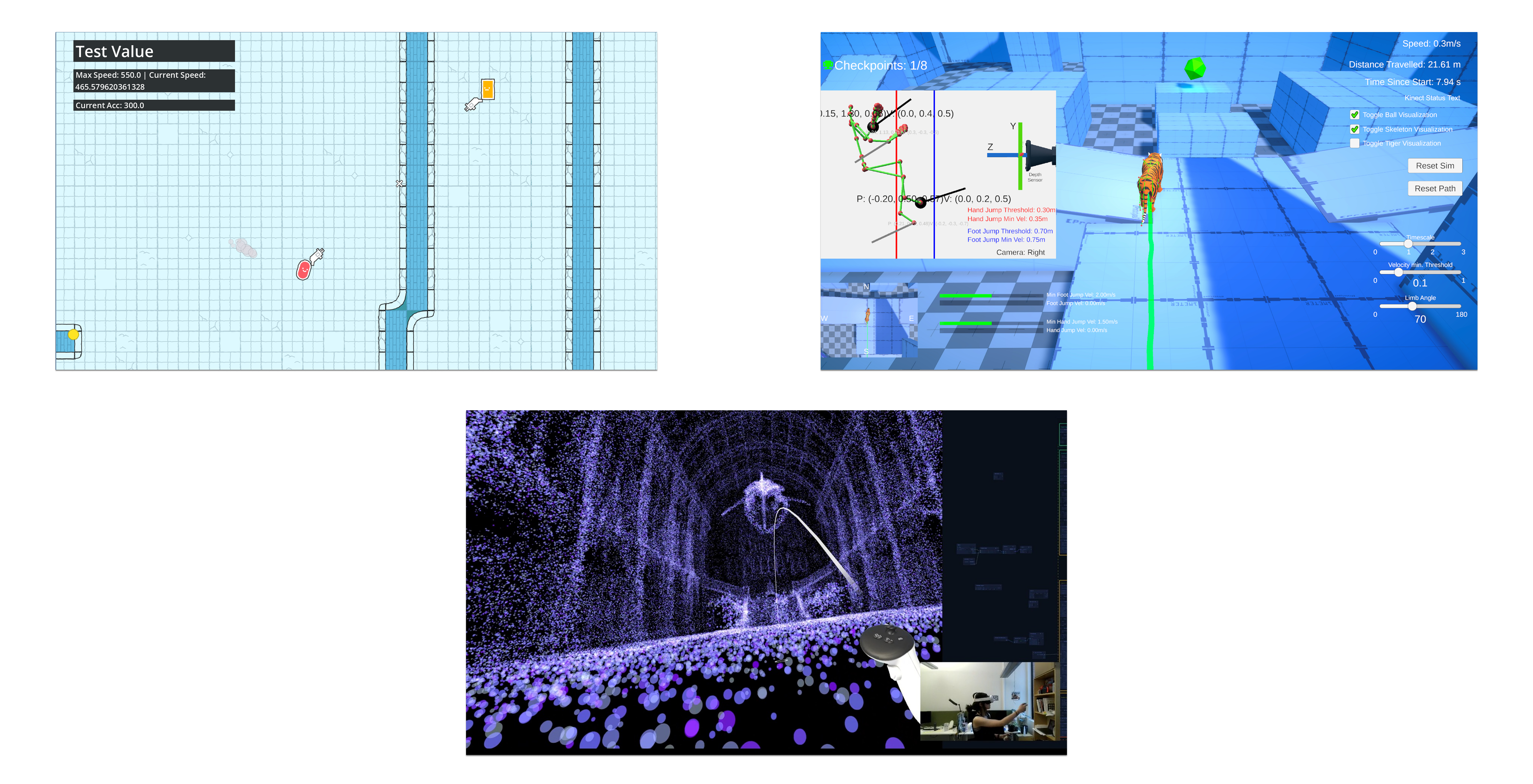}
    \caption{Screenshots from the design projects undertaken by three researchers. Top-left: R1, Top-right: R3, Bottom-center: R2.}
    \Description[3 screenshots]{The screenshot from R1's project shows a 2D action game. R2's project shows a VR project in a point cloud based environment R2's camera feed is also present of the bottom right when them using a VR headset. This was setup on OBS Studio. R3's project's project shows a 3D tiger model in a greybox level. There is a sub-screen that is displaying human joint data.}
    \label{fig: project_screenshots}
\end{figure}

\subsubsection{Framing Digital Heritage as Dynamic, Multisensory Inhabitation}

R2's project was a Virtual Reality experience, exploring how sound heritage (source, time, place, context, acoustic data) could be mapped into a VR space. Their design goal was to use RDA to document where their experiments with sound, interaction, and embodiment were drifting, converging, or unexpectedly opening new pathways. They found RDA most useful during the later “polish” and “UX design” phases, where reflection on experiential qualities became central. The functional prototyping was not prioritised for journaling, since solving such problems was not regarded as part of their design goals. The tool had not been built with VR in mind, so the initial setup needed some modding to make the tool compatible, for example, disabling the VR rig when triggering the post test reflection as it caused some conflicts with the standard UI the tool utilized.

A key moment in their project was when they received critical feedback on their project, which made them doubt their design process. During that moment of distress, it led them to doubt even their research orientation. Revisiting the journals helped them retrace past rationale, not to resolve the doubts immediately, but to better understand their trajectory. This moment also made them understand how confidence, validation, and criticism can directly affect design trajectories and the researcher’s well-being. During analysis, they found the chronological compilations insightful, but inadequate tagging throughout the process, due to still-developing project goals, made thematic compilations challenging to study. They are open to using RDA in future academic and non-academic projects, where reflection and creativity are central to a project. They note their biggest need for the tool would be improved tagging and tag-organisation.

R2 recorded 190 reflections over 25 sessions. The total data corpus size was 27.2 GBs. Note that there were a couple of instances when R2 accidentally quit a test session without using the custom key bind and that left OBS recording in the background till R2 noticed and shut down the recording manually. Those accidental runs were filtered from the corpus size calculation.

\subsubsection{The Technical Challenge of Human-Motion Driven Quadrupedal Movement}

R3's engaged in the development of an exercise game where the player controlled a quadrupedal character (a tiger) through human body movements, tracked using a Microsoft Kinect\footnote{https://azure.microsoft.com/en-us/products/kinect-dk}. They used RDA to organize their prototyping process and utilize the journaled challenges and hurdles as learning opportunities. Their reflection process leaned on a more technical side, where they would write a design goal in the pre-test comment and then engage in experimentation and problem-solving during the test. If a solution was reached, they would write the details as the post-test comment. Otherwise, they would plan upcoming steps necessary to solve the issue. During initial RDA engagement, they became more cautious of performing tests, making sure that everything was programmed correctly in order to avoid redundant journals. But this feeling dissipated with experience, and prolonged RDA engagement felt to improve overall project focus.

Eventually, the logs became part of their design process, serving both as a knowledge base that informed future decisions and as a design reference. Referring back to previous entries helped them track progress, verify fixes, do sanity checks, and check cause–and–effect relationships. They are optimistic about using RDA in future academic projects, noting that RDA can be used in quantitative, qualitative and mixed-methods contexts. They speculate future uses of the tool, such as automating the generation of a game design document from journal data or using the tool to journal AI-assisted game development. They are also excited to explore RDA in non-academic contexts, hoping to use RDA in experimental projects where they would step out of their comfort zone and learn something new, like learning a new game engine. 

R3 recorded 1110 reflections over 35 sessions. The total data corpus size was 16.2 GBs.

\subsection{Findings from Self Usage}

\subsubsection{Designer-Routine Compromise}
\label{sec:4.1}

The designers each had established routines, habits, and workflows, some of which were based on their preferences, while others were based on the details of the project itself. The ways in which RDA interacted with or conflicted with these design routines led to the designers' emergent behaviours and attitudes. Some routines are more general, like a cultural practice shared by all three designers, for example, rapid prototyping. However, other routines (or even the fine details of the general routines) can be quite unique and specific for each designer. This leads to the user experience of RDA being characteristically subjective for each designer. As the tool was initially developed and driven by the needs of R1 and their respective project, when R2 and R3 adopted the tool, they faced unique challenges, which acted as points of frustration, especially in the beginning.

\begin{enumerate}[topsep=0pt,itemsep=-1ex,partopsep=1ex,parsep=1ex]
    \item {\textit{Tool Modding}}: The friction brought about by these challenges, though frustrating, also had a relatively positive result, wherein the designers inculcated a tinkering mindset. They would keep brainstorming ways in which the tool can be modded to adjust their specific needs and styles, and where needed, they would make changes and personalize the tool. In this sense, the flexible nature of the tool promoted a reactive development style where the designer and the RDA tool would occasionally have a "conversation".

\begin{quote}
    \textit{"It’s good for a VR designer to record the movement in real life as well I believe, and it’s much easier to record in OBS. I like it!"} - R2's diary entry when they incorporated video recordings in addition to gameplay recordings through OBS.
\end{quote}

\item{\textit{Designer Modding}}: This conversation between the designer and RDA wasn't one-sided. As the designers moulded the tool to suit their needs, incorporating RDA into the workflow also required designers to adapt. Some of these adaptations were the result of an active effort from the designer, for example, being more reflective in the design process and learning new muscle memory. However, other adaptations were an emergent effect of engaging with RDA, for example, clearer focus and improved ability to verbalize tacit knowledge. It should be noted that this habit adjustment is still a cost and can affect development speeds.

\begin{quote}
    \textit{ "My workflow changes might also involve modifications to the tool to better fit the project requirements beforehand, so that I can save time overhead in the long run. The tool and the project should ideally have a symbiotic and mutually beneficial relationship that would not cause friction with each other." }- R3 when speculating on how/whether they would use their tools in future projects.
\end{quote}

\item{\textit{Value in Reflection Itself}}: As RDA became a more regular and natural part of the workflow, the designers also grew better at explicating their tacit knowledge and decisions. Thus, the moments of reflection themselves started to become points where design insights were generated. The reflection-comment moments became avenues for critical thinking, problem solving, and refining one's design process.

\begin{quote}
    \textit{"Using the tool also feels like a unique learning experience as the critical perspective makes you question your design processes in ways you would otherwise not be likely to, and in that sense you can explore and learn a lot there."} - R1 when reflecting on the biggest benefits of the tool use.
\end{quote}
\end{enumerate}

\subsubsection{Designer-Researcher Persona Consolidation}
\label{sec:4.2}

The research through design process requires the researcher to actively participate from both a researcher and designer's perspective. In the RDA process, the reflection is placed close to the moment of design materialization. During reflection, it is the researcher persona that needs to be active, while before and during design materialization, the designer persona needs to be present. This frequent gear-switching between the designer and researcher can lead to a unique developer experience.

\begin{enumerate}[topsep=0pt,itemsep=-1ex,partopsep=1ex,parsep=1ex]

\item{\textit{A Fresh Attitude and Approach Required}}: As mentioned in an earlier theme, designers can have their own set of processes and routines. Similarly, research and critical thinking are accompanied by their own set of practices, and these two may cause conflicts. Being in the "flow state" during design may interfere with the reflection process, leading to both disruption of flow and poor critical thinking. However these interruptions were not always seen as completely negative, as the subtle friction can actually allow for a smoother shift between the designer and researcher mindset. One perspective to take is that the RDA process calls for its own routines and rituals, accepting that this process is distinct and should be approached as such. As discussed with the Designer and Tool Modding, the designers eventually settled into a practice that worked best for them, so it is also a matter of experience and familiarity.

\begin{quote}
    \textit{"Perhaps the most important benefit was that [the RDA tool] introduced structured pauses in my workflow. Even though these pauses sometimes broke my flow, they also prompted me to articulate decisions that I might otherwise have left implicit."} - R2 reflecting on their overall experience with the RDA tool.
\end{quote}

\item{\textit{Selective Tool Usage}}: Although there was no explicit instruction on how often designers should use the tool, they all started their projects using the tool at every test step. Eventually, each of them reached the conclusion that there were some tests, which did not benefit from reflection and where skipping the journaling would ease the process, for example, during bug-fixing or rapid prototyping. The designers were able to identify which sections of the project were richer in their desired data and focused testing in those sections. These journaling omissions were not always intentional or ideal, sometimes they would arise from external factors such as deadline pressure, where the time added by journaling would actually start to matter and then a designer might choose to skip reflection in favour of project completion. 

Following selective tool usage, not every test was recorded, and thus some portion of data was lost. But on the contrary, if every small detail is hypothetically recorded, depending on the span of the project, the data volume can bloat making later analysis challenging. This is especially true if good tagging strategies were only developed later in the process.  This subtle mental conflict can stay present at the back of the designer's mind and have influence on their process negatively. To mitigate this, the designer has to find the balance between data volume and data loss that they feel comfortable with.

\begin{quote}
\textit{"Stopped using the tool for now since it seems to obstruct and slow down the rapid testing process. Maybe I will turn it back on once the current state of the project is more stable / requires less frequent testing."} - a note by R3 in their design diary.
\end{quote}

\end{enumerate}

\subsubsection{Mirror Effect of RDA}
\label{sec:4.3}

As the designers got more accustomed with RDA and it became integrated into their project, it was able to actively capture designers' decisions and patterns. In doing so, it started to act as a mirror that revealed the designers' process otherwise tends to be messy but invisible. This transparency started to trigger self-consciousness to varying degrees. Each designer had a different reaction to this, which in turn influenced their projects.

\begin{enumerate}[topsep=0pt,itemsep=-1ex,partopsep=1ex,parsep=1ex]
\item{\textit{Effects of Self-Consciousness}}: The recording of the creative and technical practice creates the possibility of being judged by others. But what can be more immediate is being judged by oneself. This feeling and judgment can then have noticeable effects on the journaling style and process of the designer. The designer may want to avoid recording failures and curating data, feeding into the aforementioned selective tool use. The messy and chaotic design process, which can capture important insights, can thus remain invisible. This can also lead to designers performing value judgements, asking if certain tests are even worth recording. This self-consciousness was relatively subtle for each designer, and two designers noted explicitly that continuous engagement with RDA and naturalization of the process helped alleviate this feeling.

\begin{quote}
    \textit{" ... to make the entire design process, especially moments of frustration and failure [visible] feel a little challenging. But I think it’s something that I came to terms with and it didn’t really bother me much."} - R1 reflecting whether the tool made them feel monitored or self-consciousness.
\end{quote}

\item{\textit{A Reflection of the Designer}}: On the flip side, the use of the tool and the collected data was able to act as a reflection of designer themselves, which would otherwise be hard to capture. RDA captures habits, patterns and a shifting mindset that can then be reflected on. This can act as natural learning moments where the improvements can be discovered, the research process can be revised or just self-understanding improved. The designer's emotional responses can also be captured, noting how they react and process various moments in the design process on an affective level.

\begin{quote}
    \textit{"[The RDA tool] might not only capture design decisions but also provide a lens into the affective ecology of RtD practice — where anxiety, doubt, or encouragement are not peripheral, but integral to how design knowledge is produced." }- R2 when reflecting on what the RDA tool helped them explore outside of their specific design goals.
    
\end{quote}
\end{enumerate}

\begin{figure}
    \centering
    \includegraphics[width = \linewidth]{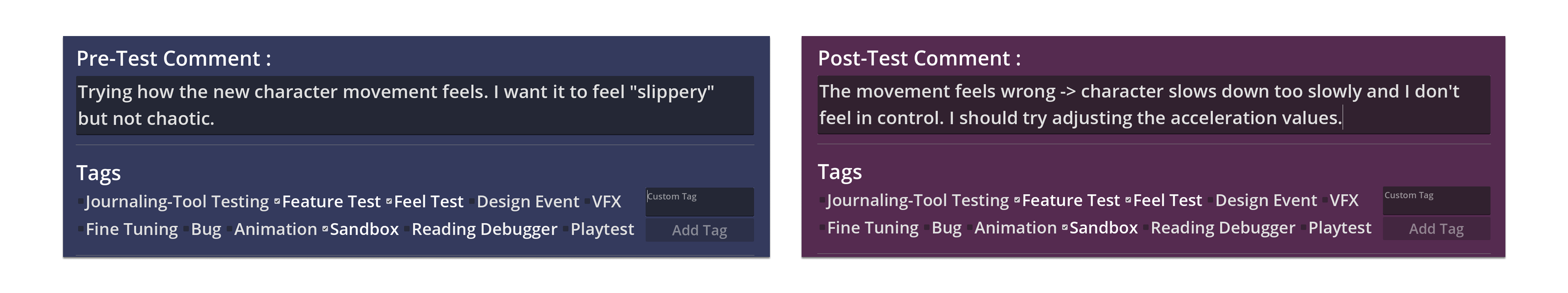}
    \caption{An illustrative example of the pre-test and post-test reflection journaling. In the pre-test comment, the designer reflects on their expectations about the mechanic they just designed. In the post-test reflection, they compare the play-experience with their expectation and speculate on the next design step.}
    \Description[Two screenshots of the post test and pre test UI]{On the left is the pre-test UI, it is a deep blue in color. It contains the comment, "Trying how the new character movement feels. I want it to feel 'slippery' but not chaotic" and has the tags Feature Test, Feel Test and Sandbox Selected. On the right is the post-test UI, it is a deep magenta in color. It contains the comment, "The movement feels wrong. Character slows down too slowly and I don't feel in control. I should try adjusting the acceleration values.". The tags are unchanged.}
    \label{fig:reflection_example}
\end{figure}

\subsection{Evaluation}
\label{sec:4.4}

We compare our experiences against the core requirements of RDA that were established in Section~\ref{sec:4.2_requirements}. 


    \begin{enumerate} [topsep=0pt,itemsep=-1ex,partopsep=1ex,parsep=1ex]
        \item \textit{Capturing micro-design reflections}: The designers agree that the tool offered the capability to capture micro-design reflections in a well-organized fashion, noting it as one of the biggest advantages of using the tool. R1 noted that finer details of sub-design tasks that happen outside the playtest like sound and visual effects, can be missed. R3 points out that there is room for improvement, stating that the reflection's worth depends on how regular and detailed the reflections are. They also hinted at the possibility of logging screenshots for improved clarity and communication. 
        \item \textit{Track holistic evolution through automatic video recordings}: The designers found that the videos were consistently recorded, and the compilation allowed to visually observe the chronological and thematic growth of the project. R2 notes that tracking thematic evolution depended heavily on good tagging schemes. R3 adds that linking these logs to version control commits might give an even more detailed sense of the long-term project growth.
        \item \textit{Minimizing intrusion to the design process}: The designers experienced onboarding friction, wherein the most intrusive moments tended to occur when first engaging with RDA. With continuous use and developing a personalized workflow, RDA felt more naturalized, with each designer noting that the tool use eventually became intuitive. 
        \item \textit{Processing raw data to be digestible and analyzable}: The designers felt that the compiled data improved readability and aided in analysis. Being able to scrub through all the tests from a day in a single file was especially helpful. R1 found the data easy to use on qualitative data analysis software like Atlas.ti. 
    \end{enumerate}

Overall, the authors agreed that the tool fulfilled the core requirements well and added value to their design process. It still posed challenges given its nature as a prototype and we highlight some the prominent issues below: 

\begin{itemize}
    \item RDA has only been tested at the individual level so far. Game development projects are seldom done in isolation, and RDA is yet to be scaled and tested with multiple designers working on a single project.
    \item The custom keybind poses a challenging new habit to learn. The more used to game development one is, the more likely it is that they will accidentally quit the test without using the custom keybind. R3 proposes a solution of adding a UI button throughout the test which can be used to end the test instead of the custom keybind 
    \item The current tool is challenging to work with across multiple devices. Although the json logs could be maintained over version control, the relatively large videos are hard to sync. R1 developed a habit of carrying the logs on a flash drive to sync between their work and home offices.
    \item In the same vein, depending on the length and quality of recordings and the visual fidelity of the project, the data size can add up. Each test log was between 30 seconds to three minutes long, but occasional outliers could be upwards of five minutes.
    \item Modding the tool and making changes to the tool is programming intensive.
    \item Using the RDA tool with other tools such as Github Desktop\footnote{https://github.com/apps/desktop} can make one's development setup uncomfortably complex.
    \item RDA can interfere with certain development paradigms like rapid prototyping.
    \item Tool being implemented in-engine means that it can conflict with in-game elements like game UI. 
\end{itemize}
\section{Discussion}


\subsection{An Addition to the Research-through-Design Toolkit}

Our work contributes to furthering the discourse on research through design within HCI research. RDA  provides opportunities for both explicating tacit knowledge while also engaging in a conversation with the design artifact and process, which is considered a core feature of RtD by Godin and Zahedi~\cite{godin_aspects_2014}.  It further allows viewing the interplay of design and research from a fresh perspective. We do not claim to have solved the tensions or achieved complete consolidation, but rather we add to the discussions by various other authors. As Basballe and Halskov frame it,\textit{ "... better understanding of the dynamics affecting the interplay of design and research interests may provide insight that may guide future Research through Design processes, and improve the benefits for both research and design"}~\cite{basballe_dynamics_2012}. We further posit that RDA can be employed for the specific study of the interplay of design and research, advancing the study of coupling, interweaving and decoupling of design and research that Basballe and Halskov first explored~\cite{basballe_dynamics_2012}.

Focusing on research through design in games, RDA provides scaffolding for the collection and analysis of design data in a vein similar to the Method for Design Materialization (MDM) process~\cite{khaled_generative_2023}. Though parallel approaches, RDA focuses more on visual capture of design recording over code changes, and reflecting during design rather than outside of it. The MDM process, in that sense, minimizes any intrusion to the design process, at the cost of reflection granularity. But that is again very much dependent on designer practice, and more technically fluent designers can engage with version control more often than with in-game tests. We believe that MDM, RDA and other available research through design approaches to be complementary, and it is up to designers to choose which works best with their practice. It could be one or the other, a hybrid approach or a new approach that the current design documentation tools fail to provide and capture. These approaches are well imagined as modular building blocks, for example, R1 experimented with the use of temporal vocabulary~\cite{oogjes_temporal_2024} in their process, and it worked quite smoothly with the RDA process.

\subsection{Value of Finding Points of Design Actualization}

One of the key characteristics of RDA is the identification and utilization of moments of design actualization as points of reflection. There are already many established approaches to reflection in design studies. One common method is to reflect when a moment of surprise or insight hits the designer. RDA, to some degree, goes against this practice by not waiting for insight to strike, but rather placing reflection moments at design-critical moments. which is supported by authors like Sengers et al.~\cite{sengers_reflective_2005}. But at the same time, they also warn against \textit{"a literal codification of reflection-in-action, for example pop-up windows that suggest ‘now would be a good time to think about what is happening'...’’}~\cite{sengers_reflective_2005}. RDA treads a fine line between these two camps and depends on designer preferences. 

Finding the right amount of interaction with the tool and identifying spaces to consciously skip reflections can help avoid this codification. Additionally, the process of this structured reflection and insight-based reflection need not be mutually exclusive. In our own practices, each researcher maintained some other form of journals along with the RDA tool, be it a diary or version control logs. We would encourage designers from all spheres to at least experiment with identifying what design actualization could mean in their processes and how spotlighting it can influence the overall design process.

\subsection{Possibilities and Future Explorations}

We are also intrigued by the possibility of employing RDA as a learning and teaching tool. Research through design has already been proven to be well-suited for teaching interaction design~\cite{hansen_teaching_2018} and it can similarly be extended to game design learning, which is already very hands-on. Students tend to learn their craft through practice and observation, and adding a layer of critical reflection can then bolster their understanding of their own tacit practices and reasoning. RDA may also act as a structured template for students approaching research-through-design for the first time~\cite{hansen_teaching_2018}.

Another opportunity to explore is to study the dialogue between designers and players. Though we briefly touched upon it during our testing, we would like to further explore how playtesting with players influences designer actions and attitudes. This helps connect the two sides of game studies and help bring the player and design experience in the same space. In general, it seems RDA has great potential for not just studying processes and artifacts, but also designers. Exploration of designer behaviours, emotional journeys, and learning experiences all present avenues for future research.

RDA provides a new option to bolster analytical rigour for academics who want to participate in game design and game designers who want to engage in their craft within academia. We hope RDA encourages more designers to make games that don't need to be made~\cite{bates_making_2025}, and engage in cultural critique with games in line with critical design, reflective design and critical making~\cite{bardzell_what_2013, ratto_critical_2011, sengers_reflective_2005}. Extending this idea, we also hope to support engagements with ludic design~\cite{gaver_drift_2004}, where curiosity and exploration are encouraged through the use of the designed artifact over concrete goal-based design. Failures are another facet of design rarely explored in HCI research. Howell at al. have called attention to this by challenging success narratives and establishing value to studying failure and challenging design moment~\cite{howell_cracks_2021}. \textit{“We examined lingering feelings of discomfort, frustration, or guilt with the intuition that something of value lay hidden beside the stories of normative success. Through this, we articulated less glossy but ultimately richer stories of our past projects and ourselves as design researchers”}~\cite{howell_cracks_2021}. This narrative echoes our own experiences which we were able to capture, process and understand through engagement with RDA. Similar calls for understanding "janky" experiences in games have also been made~\cite{bennett_jank_2023}. 

RDA can potentially be applicable in collaborations between industry and academia. In their game-designer interview study, Denisova et al.~\cite{denisova_whatever_2021} point out that retrospective accounts of game creation may not accurately reflect the day-to-day design experience. They call for longitudical studies that follow the journeys of game projects, which is a gap that RDA is well suited to cover. Though an exciting prospect, one should also note that the production pace and stakeholders in industry tend to be starkly different from academia. The friction and the time investment required by RDA might be considered too heavy, but this can only be understood through discussion, experimentation and collaboration.



\section{Limitations and Future Work}




Adopting an autobiographical design approach allowed us to utilize our needs, experiences, and personal understanding to perform a first exploration and construction of RDA in a  longitudinal and flexible fashion. Yet there are limitations to the outcomes of autobiographical design. To better understand how RDA works with a wide range of designers and design processes, more designers and researchers need to experiment with the it. Although it may be challenging to recruit external participants with game design and research backgrounds for long-term studies like ours, integrating RDA in academic game jams in game design courses might be suitable if prioritizing sample size over study duration.  Additionally, studies comparing the engagement with and the data collected using multiple approaches such as RDA, MDM, and annotated journals could be insightful for revealing the intricacies of each process.
\section{Conclusion}

With the evolving landscape of research through design in games, there is a growing need for tools and processes for the documentation and analysis of varied game design projects. We identified two such requirements, namely, capturing micro design decisions and recording holistic design evolution. To address this, we engaged in an autobiographical design process to develop and evaluate Reflection at Design Actualization (RDA). RDA proved capable of tackling the aforementioned requirements, while also balancing intrusiveness to the design process and data digestibility. Being closely threaded into the design process itself, the use of RDA led to a unique designer experience, which in the end inculcated rich and critical design evaluation and understanding. We also shared and discussed the unique challenges of the RDA tool, and possible ways to mitigate them.  We hope that RDA can be an addition to the research through design toolkit within games and HCI, allowing for a better understanding of games, the game design and game designers.

\section{Supplementary Material}


Readers can find the RDA took files for Godot and Unity, along with the Python compilation script with their respective setup and usage instructions here: \url{https://github.com/Prabhav2B/Reflection-at-Design-Actualization-Tool}


\renewcommand{\bibnumfmt}[1]{[#1]}%
\bibliographystyle{ACM-Reference-Format}
\bibliography{references}

\end{document}